\documentstyle[12pt]{article}
\newcounter{appendixc}
\newcounter{subappendixc}[appendixc]
\newcounter{subsubappendixc}[subappendixc]

\renewcommand{\appendix}[1] {\vspace*{0.6cm}
        \refstepcounter{appendixc}
        \setcounter{figure}{0}
        \setcounter{table}{0}
        \setcounter{equation}{0}
        \renewcommand{\thefigure}{\Alph{appendixc}.\arabic{figure}}
        \renewcommand{\thetable}{\Alph{appendixc}.\arabic{table}}
        \renewcommand{\theappendixc}{\Alph{appendixc}}
        \renewcommand{\theequation}{\Alph{appendixc}.\arabic{equation}}
        \noindent{\bf Appendix \theappendixc #1}\par\vspace*{0.4cm}}

\begin{document}
\begin{titlepage}
\baselineskip=0.25in
\begin{flushright}
AMES-HET-97-1\\
February 1997
\end{flushright}
\vspace{0.1in}
\begin{center}
{\Large Dimension-six CP-conserving operators of the third-family quarks\\      
        and their effects on collider observables }
\vspace{.2in}

         K.Whisnant$^a$, Jin Min Yang$^{a,b,c}$, Bing-Lin Young$^{a,b}$,
         and X.Zhang$^d$
\vspace{.2in}

\it
$^a$     Department of Physics and Astronomy, Iowa State University,\\
         Ames, Iowa 50011, USA\\
$^b$     International Institute of Theoretical and Applied Physics,\\
         Iowa State University, Ames, Iowa 50011, USA\\
$^c$     Physics Department, Henan Normal University,\\
         Xin Xiang, Henan 453002, China\\
$^d$     Institute of High Energy Physics, Academia Sinica, \\
         Beijing 100039, China
\end{center}
\vspace{.4in}
\rm
            \begin{center} ABSTRACT\end{center}

We list all possible dimension-six CP-conserving
$SU_c(3)\times SU_L(2)\times U_Y(1)$ invariant operators involving the 
third-family quarks which could be generated by new physics at a higher scale. 
Expressions for these operators after electroweak gauge symmetry breaking 
and the induced effective couplings $Wt\bar b$, $Xb\bar b$
and $Xt\bar t$ $(X=Z,\gamma,g,H)$ are presented .
Analytic expressions for the tree level contributions of all these operators 
to the observables $R_b$ and $A^b_{FB}$ at LEP I, 
$\sigma(e^+e^-\rightarrow b\bar b)$ and $A^b_{FB}$ at 
LEP II, $\sigma(e^+e^-\rightarrow t\bar t)$ and $A_{FB}^t$ at the NLC, as well 
as $\sigma(p\bar p\rightarrow t\bar b+X)$ at the Tevatron upgrade, are provided.
The effects of these operators on different electroweak observables are 
discussed and numerical examples presented. Numerical analyses
show that in the coupling region allowed by $R_b$ and $A^b_{FB}$ at LEP I, 
some of the new physics operators can still have significant contributions
at LEP II, the Tevatron and the NLC.  
\end{titlepage}
\eject

\baselineskip=0.30in
\begin{center}{\Large 1. Introduction}\end{center}

The Standard Model (SM) has been very successful phenomenologically[1].
The discovery of the top quark[2] fulfilled the long 
anticipated
completion of the fermion sector of the SM. Despite its success, the SM
is still believed to be a theory effective at the electroweak scale and
that some new physics must exist at higher energy regimes. 
The exceedingly large mass of the top quark further strengthens this belief.
Collider
experiments have been used to search for the new particles predicted by
various new physics models, but no direct signal of new
particles has been observed. So, if new physics indeed exists above the
electroweak scale, it is very likely that the only observable effects 
at energies not too far
above the SM energy scale could be in the form of new interactions
affecting the couplings of the third-family quarks, and the untested sectors of
the Higgs and
gauge bosons. In this spirit, the new physics effects can be expressed
as non-standard terms in an effective Lagrangian describing the
interactions among third-family quarks, the Higgs and gauge
bosons with a form like, before the electroweak symmetry breaking,
\begin{equation}
{\cal L}_{eff}={\cal L}_0+\frac{1}{\Lambda^2}\sum_i C_i O_i
                         +O(\frac{1}{\Lambda^4})
\end{equation}
where ${\cal L}_0$ is the SM Lagrangian, $\Lambda$ is the new physics
scale, $O_i$ are dimension-six operators which are
$SU_c(3)\times SU_L(2)\times U_Y(1)$ invariant
before the electroweak symmetry break-down, and $C_i$ are
constants which represent the coupling strengths of $O_i$.  
The expansion in Eq.(1) was first
discussed in Ref. [3]. 

Further classification of the operators $O_i$ has been made more recently.
The CP-conserving operators involving only
weak bosons were classified and phenomenological implications discussed
in Ref.~[4]. The corresponding operators involving the third-family quarks 
were enumerated in
Ref.~[5]. In these earlier works [3,4,5] the field equations of all particles
were used to reduce the number of operators in Eq.(1). 
The phenomenology of some of these CP-conserving operators were discussed
in Refs.[6-8]. More recently the operators were reconsidered without using
the field equations of the gauge bosons[9].  
In this article, we again focus on the set of operators 
involving the third-family quarks.
We use the most recent LEP I data involving the $b\bar b$ final state to
constrain some of the coefficients $C_i$, assuming the simple 
situation that cancellation among different operators does not take place.
We identify the operators which can potentially have significant effects
on the standard model predictions at higher energies in LEP II, the NLC and 
the Tevatron.

This paper is organized as follows. 
In Sec.~2 we again list all possible operators in Eq.(1).
The expressions for these operators after electroweak gauge symmetry breaking 
and the induced effective couplings $Wt\bar b$, $Xb\bar b$
and $Xt\bar t$ $(X=Z,\gamma,g,H)$ are presented in Appendices A and B.
In Sec.~3 we give analytic expressions for the contributions of  
these operators to the observables $R_b$ and $A^b_{FB}$ at LEP I, 
$\sigma(e^+e^-\rightarrow b\bar b)$ and $A^b_{FB}$ at 
LEP II, $\sigma(e^+e^-\rightarrow t\bar t)$ and $A_{FB}^t$ at the NLC as
well as  $\sigma(p\bar p\rightarrow t\bar b+X)$ at the Tevatron.
In Sec.~4 we determine which collider observables are affected by each 
operator. In Sec.~5 we analyze the operators which
affect $R_b$ and $A_{FB}^b$ at LEP I and determine how much they affect 
future electroweak collider observables, subject to current constraints.
Finally, in Sec.6 we conclude with some discussion and a summary.
\vspace{1cm}

\begin{center} {\Large 2. CP-conserving gauge invariant operators}
\end{center}

We follow the conventional notation which is listed below 
\vspace{.5cm}
\begin{eqnarray*}
 \begin{array}{lll}
{\rm left-handed~ third-family~ doublet:}&~~~~& 
 q_L=\left (\begin{array}{l}t_L\\ b_L\end{array}\right),\\
 &~~~~&  \\
{\rm right-handed~ top,~bottom~ quarks:}&~~~~&t_R,~~b_R,\\
 &~~~~&  \\
{\rm Higgs~ boson~ doublet:}&~~~~&\Phi,~~\widetilde\Phi=i\sigma_2\Phi^*,\\
 &~~~~&  \\
{\rm gluon~ fields:}&~~~~&G^A_{\mu},~~A=1\cdots 8,\\
             &~~~~& G^A_{\mu\nu}=\partial_{\mu}G^A_{\nu}
                     -\partial_{\nu}G^A_{\mu}        
                     +g_sf_{ABC}G^B_{\mu}G^C_{\nu},\\
             &~~~~& G_{\mu}=T^AG^A_{\mu},~~ G_{\mu\nu}=T^AG^A_{\mu\nu}, 
                     ~~T^A=\lambda^A/2,\\
 &~~~~&  \\
SU_L(2)~{\rm gauge~ fields:}&~~~~&W^I_{\mu},~~ I=1\cdots 3,\\
                       &~~~~&W^I_{\mu\nu}=\partial_{\mu}W^I_{\nu}
                     -\partial_{\nu}W^I_{\mu}        
                     +g_2\epsilon_{IJK}W^J_{\mu}W^K_{\nu},\\
             &~~~~& W_{\mu}=\tau^IW^I_{\mu},~~ W_{\mu\nu}=\tau^IW^I_{\mu\nu}, 
                     ~~\tau^I=\sigma^I/2,\\
 &~~~~&  \\
U_Y(1)~{\rm gauge~ field:} &~~~~& B_{\mu},\\
                       &~~~~& B_{\mu\nu}=\partial_{\mu}B_{\nu}
                     -\partial_{\nu}B_{\mu}.        
\end{array}
\end{eqnarray*}
\vspace{.5cm}

In order to justify the forms of operators that we will use,
let us elaborate on the origin of the new physics  
that has been touched upon in Ref.[6].
We assume that the new physics in the quark sector resides in the third quark 
family. Before the electroweak symmetry breaking all dimension-6 
operators containing the third family quarks are possible. Although new physics
may also occur in the gauge boson and Higgs sectors, or give rise to four-quark
 operators involving the third family quarks, for the purpose of testing new 
physics in the immediate and near future, such operators will be ignored.
Therefore, the operators we are interested in  are those containing quarks
and gauge and Higgs bosons.

To restrict ourselves to new physics of the lowest order, in both the standard
model coupling and the power of $1/\Lambda^2$, we consider only tree diagrams
which contain only one anomalous vertex in a given diagram. Under these 
assumptions, operators which can be related by the field equations of the 
fermions are no longer independent and the fermion equations of motion can 
be used to reduce the number of 
independent operators, as was done in Ref.[3]. However, we have to
be careful in applying the field equations of the bosons.
Under the assumption of the new physics origin as given above, the
equations of motion of the gauge bosons can not be used when first
writing down the operators in Eq.(1). This is because the field equations
of the gauge bosons will lead to four-fermion operators containing third family
quarks and light fermions. Then naively applying the criterion of ignoring
all four-fermion operators, which are observable in the existing colliders,
e.g. $e^++e^-\rightarrow b+\bar b$,
would discard these operators which originate from 
new physics different from that of the four-fermion operators discarded 
initially. However,
the equation of motion of the Higgs field can be used since the light fermions
resulting from the Higgs field equations are proportional to $m_l/m_W$, where
$m_l$ is the mass of the light fermions concerned. 

We should also remark that no field equation can  be used in the case of 
loop diagrams or when new
physics couplings appear more than once in a tree diagram. 
In the latter case, dimension-8 operators may also have to be included.
This means that extending the effective Lagrangian approach to
dimension-8 operators will greatly increase the number of
independent operators.

Now we list all possible dimension-six CP-conserving
$SU_c(3)\times SU_L(2)\times U_Y(1)$ invariant independent
operators involving third-family quarks but no four-fermion operators
under the qualification described above.

(1) Class 1 (containing $t_R$ ){\footnote{ 
It is straight forward to show that the last two operators $O_{tG}$ and 
$O_{tB}$ can be recast into simple forms, e.g. $O_{tG}=-\bar t\gamma^{\mu}
T^a t D^{\nu}G_{\mu\nu}$, etc.}}
\begin{eqnarray}
O_{t1}&=&(\Phi^{\dagger}\Phi-\frac{v^2}{2})\left [\bar q_L t_R\widetilde\Phi
         +\widetilde\Phi^{\dagger} \bar t_R q_L\right ],\\
O_{t2}&=&i\left [\Phi^{\dagger}D_{\mu}\Phi
         -(D_{\mu}\Phi)^{\dagger}\Phi\right ]\bar t_R \gamma^{\mu}t_R,\\
O_{t3}&=&i\left [(\widetilde\Phi^{\dagger}D_{\mu}\Phi)(\bar t_R \gamma^{\mu}b_R)
         -(D_{\mu}\Phi)^{\dagger}\widetilde\Phi(\bar b_R \gamma^{\mu}t_R)
       \right ],\\
O_{Dt}&=&(\bar q_L D_{\mu} t_R) D^{\mu}\widetilde\Phi
         +(D^{\mu}\widetilde\Phi)^{\dagger}(\overline{D_{\mu}t_R}q_L),\\
O_{tW\Phi}&=&\left [(\bar q_L \sigma^{\mu\nu}\tau^I t_R) \widetilde\Phi
         +\widetilde\Phi^{\dagger}(\bar t_R \sigma^{\mu\nu}\tau^I q_L)\right ]
          W^I_{\mu\nu},\\
O_{tB\Phi}&=&\left [(\bar q_L \sigma^{\mu\nu} t_R) \widetilde\Phi
         +\widetilde\Phi^{\dagger}(\bar t_R \sigma^{\mu\nu} q_L)\right ]
          B_{\mu\nu},\\
O_{tG\Phi}&=&\left [(\bar q_L \sigma^{\mu\nu}T^A t_R) \widetilde\Phi
         +\widetilde\Phi^{\dagger}(\bar t_R \sigma^{\mu\nu}T^A q_L)\right ]
          G^A_{\mu\nu},\\
O_{tB}&=&\left [\bar t_R \gamma^{\mu} D^{\nu}t_R
         +\overline{D^{\nu}t_R} \gamma^{\mu} t_R\right ]
          B_{\mu\nu},\\
O_{tG}&=&\left [\bar t_R\gamma^{\mu}T^A D^{\nu}t_R
         +\overline{D^{\nu}t_R} \gamma^{\mu}T^A t_R\right ]
          G^A_{\mu\nu},
\end{eqnarray}
(2) Class 2 ( not containing $t_R$){\footnote{ 
It is straight forward to show that the first five operators, $O_{qG}$
$O_{qW}$,$O_{qB}$,$O_{bG}$ and $O_{bB}$, 
can be recast into simple forms, e.g. $O_{qG}=-\bar q_L\gamma^{\mu}
T^a q_L D^{\nu}G_{\mu\nu}$, etc.}}
\begin{eqnarray}
O_{qG}&=&\left [\bar q_L \gamma^{\mu}T^A D^{\nu}q_L
         +\overline{D^{\nu}q_L} \gamma^{\mu}T^A q_L\right ]
          G^A_{\mu\nu},\\
O_{qW}&=&\left [\bar q_L \gamma^{\mu}\tau^I D^{\nu}q_L
         +\overline{D^{\nu}q_L} \gamma^{\mu}\tau^I q_L\right ]
          W^I_{\mu\nu},\\
O_{qB}&=&\left [\bar q_L \gamma^{\mu} D^{\nu}q_L
         +\overline{D^{\nu}q_L} \gamma^{\mu} q_L\right ]
          B_{\mu\nu},\\
O_{bG}&=&\left [\bar b_R \gamma^{\mu}T^A D^{\nu}b_R
         +\overline{D^{\nu}b_R} \gamma^{\mu}T^A b_R\right ]
          G^A_{\mu\nu},\\
O_{bB}&=&\left [\bar b_R \gamma^{\mu} D^{\nu}b_R
         +\overline{D^{\nu}b_R} \gamma^{\mu}b_R\right ]
          B_{\mu\nu},\\
O_{\Phi q}^{(1)}&=&i\left [\Phi^{\dagger}D_{\mu}\Phi
      -(D_{\mu}\Phi)^{\dagger}\Phi\right ]\bar q_L \gamma^{\mu}q_L,\\
O_{\Phi q}^{(3)}&=&i\left [\Phi^{\dagger}\tau^I D_{\mu}\Phi
        -(D_{\mu}\Phi)^{\dagger}\tau^I\Phi\right ]\bar q_L \gamma^{\mu}\tau^I 
       q_L,\\
O_{\Phi b}&=&i\left [\Phi^{\dagger}D_{\mu}\Phi
         -(D_{\mu}\Phi)^{\dagger}\Phi\right ]\bar b_R \gamma^{\mu}b_R,\\
O_{b1}&=&(\Phi^{\dagger}\Phi-\frac{v^2}{2})\left [\bar q_L b_R\Phi
         +\Phi^{\dagger}\bar b_R q_L\right ],\\
O_{Db}&=&(\bar q_L D_{\mu} b_R) D^{\mu}\Phi
         +(D^{\mu}\Phi)^{\dagger}(\overline{D_{\mu}b_R}q_L),\\
O_{bW\Phi}&=&\left [(\bar q_L \sigma^{\mu\nu}\tau^I b_R) \Phi
         +\Phi^{\dagger}(\bar b_R \sigma^{\mu\nu}\tau^I q_L)\right ]
          W^I_{\mu\nu},\\
O_{bB\Phi}&=&\left [(\bar q_L \sigma^{\mu\nu} b_R) \Phi
         +\Phi^{\dagger}(\bar b_R \sigma^{\mu\nu} q_L)\right ]
          B_{\mu\nu},\\
O_{bG\Phi}&=&\left [(\bar q_L \sigma^{\mu\nu}T^A b_R) \Phi
         +\Phi^{\dagger}(\bar b_R \sigma^{\mu\nu}T^A q_L)\right ]
          G^A_{\mu\nu}.
\end{eqnarray}
If we avoided using the field equations of Higgs boson and the quarks,
we would get the following additional operators\\
(3) Class 3 
\begin{eqnarray}
O_{\overline{D}t}&=&(\overline{D_{\mu}q_L} t_R) D^{\mu}\widetilde\Phi
         +(D^{\mu}\widetilde\Phi)^{\dagger}(\bar t_R D_{\mu}q_L),\\
O_{\overline{D}b}&=&(\overline{D_{\mu}q_L} b_R) D^{\mu}\Phi
         +(D^{\mu}\Phi)^{\dagger}(\bar b_R D_{\mu}q_L),\\
O_{t\widetilde B}&=&i\left [\bar t_R \gamma^{\mu} D^{\nu}t_R
         -\overline{D^{\nu}t_R} \gamma^{\mu} t_R\right ]
          \widetilde B_{\mu\nu},\\
O_{t\widetilde G}&=&i\left [\bar t_R \gamma^{\mu}T^A D^{\nu}t_R
         -\overline{D^{\nu}t_R} \gamma^{\mu}T^A t_R\right ]
          \widetilde G^A_{\mu\nu},\\
O_{b\widetilde B}&=&i\left [\bar b_R \gamma^{\mu} D^{\nu}b_R
         -\overline{D^{\nu}b_R} \gamma^{\mu}b_R\right ]
          \widetilde B_{\mu\nu},\\
O_{b\widetilde G}&=&i\left [\bar b_R \gamma^{\mu}T^A D^{\nu}b_R
         -\overline{D^{\nu}b_R} \gamma^{\mu}T^A b_R\right ]
          \widetilde G^A_{\mu\nu},\\
O_{q\widetilde B}&=&i\left [\bar q_L \gamma^{\mu} D^{\nu}q_L
         -\overline{D^{\nu}q_L} \gamma^{\mu} q_L\right ]
          \widetilde B_{\mu\nu},\\
O_{q\widetilde G}&=&i\left [\bar q_L \gamma^{\mu}T^A D^{\nu}q_L
         -\overline{D^{\nu}q_L} \gamma^{\mu}T^A q_L\right ]
          \widetilde G^A_{\mu\nu},\\
O_{q\widetilde W}&=&i\left [\bar q_L \gamma^{\mu}\tau^I D^{\nu}q_L
         -\overline{D^{\nu}q_L} \gamma^{\mu}\tau^I q_L\right ]
          \widetilde W^I_{\mu\nu},
\end{eqnarray}
where $\widetilde X_{\mu\nu}\equiv \frac{1}{2}\epsilon_{\mu\nu\lambda\rho} 
X^{\lambda\rho}$ with $X=G,B,W$ and $\epsilon_{\mu\nu\lambda\rho}$ the
anti-symmetric tensor. 
We can rewrite the above operators as follows, which will no longer be 
independent when the field equations of Higgs boson and the quarks
are used,
\begin{eqnarray}
O_{\overline{D}t}&=&-O_{Dt}-\bar q_L t_R D^2\widetilde \Phi
                           -(D^2\widetilde \Phi)^{\dagger}\bar t_R q_L,\\
O_{\overline{D}b}&=&-O_{Db}-\bar q_L b_R D^2\Phi
                           -(D^2\widetilde \Phi)^{\dagger}\bar b_R q_L,\\
O_{x\widetilde B}&=&-O_{xB}-i(\bar x_R \sigma_{\mu\nu}{\large \not}{D}x_R
-\overline{{\large \not}{D}x_R}\sigma_{\mu\nu}x_R)B^{\mu\nu},~ (x=t,b),\\
O_{x\widetilde G}&=&-O_{xG}-i(\bar x_R \sigma^{\mu\nu}T^a{\large \not}{D}x_R
-\overline{{\large \not}{D}x_R}\sigma^{\mu\nu}T^a x_R)G^a_{\mu\nu},~ (x=t,b),\\
O_{q\widetilde X}&=&O_{qX}
        +i(\bar q_L \sigma^{\mu\nu}X_{\mu\nu}{\large \not}{D}q_L
        -\overline{{\large \not}{D}xq_L}\sigma^{\mu\nu}X_{\mu\nu} q_L),
                 ~(X=B,G,W),
\end{eqnarray}

In the following analyses we will not consider the operators in Class 3 
because its operators are not independent.
Since our analyses only involve the on-shell quarks the equations of motion 
of the quarks can be applied. Because of the reasons given earlier,
the Higgs field equation can also be used.  
Then our Class 1 and Class 2 operators agree with those given in Ref.[6]. 
However, unlike Ref. [6], in $O_{t1}$ and $O_{b1}$ we subtract the
vacuum expectation value, $v^2/2$, from $\Phi^{\dagger}\Phi$,
in order to avoid additional mass terms for top and bottom quarks  
after the electroweak symmetry breaking. 

The expressions for the 
operators of Class 1 and Class 2 in the unitary gauge 
after electroweak symmetry
breaking are presented in Appendix A. From these expressions, 
one can write out the effective Lagrangian for all vertices
with two third-family fermions and a boson, specifically, $Wt\bar b$,
$Zt\bar t$, $\gamma t\bar t$, $Ht\bar t$, $gt\bar t$,  $gb\bar b$, $Zb\bar b$,
$\gamma b\bar b$ and $Hb\bar b$, whose effects are or could be reachable 
at LEP, the Tevatron upgrade and the NLC. The explicit forms of these 
effective couplings are given in Appendix B.
\vspace{1cm}

\begin{center} {\Large 3. Contributions to some collider observables} 
\end{center}

We now consider the contribution of all operators listed in Sec.~2
to the observables $R_b$ and $A^b_{FB}$ at LEP I to constrain 
the coefficients $C_i$. Then we can make predictions on their effects 
on $\sigma(e^+e^-\rightarrow b\bar b)$ and $A^b_{FB}$ at 
LEP II, $\sigma(p\bar p\rightarrow t\bar b+X)$ at the Tevatron,
$\sigma(e^+e^-\rightarrow t\bar t)$ and $A^t_{FB}$ at the NLC.
In this paper we wish to consider modifications to the electroweak
sector only, and therefore ignore measurements such as 
$\sigma(p\bar p\rightarrow t\bar t)$ which are primarily affected 
by the strong interaction.
 
Including both the SM couplings and new physics effects,  we can write
the $Zq\bar q$ and  $\gamma q\bar q$ $(q=t,b)$ vertices as
\begin{equation}\label{ver}
 \Gamma_{\mu}^{Z,\gamma}=-ieg^{Z,\gamma}\left [\gamma_{\mu}V^{Z,\gamma}_q
-\gamma_{\mu}\gamma_5 A^{Z,\gamma}_q+\frac{1}{2m_q}(p_q-p_{\bar q})_{\mu}
S^{Z,\gamma}_q\right ],
\end{equation}
where $ g^{Z}=1/(4s_Wc_W)$ with $s_W\equiv \sin\theta_W$ and 
$c_W\equiv \cos\theta_W$, $g^{\gamma}=1$,
and  $p_q$ and $p_{\bar q}$ are the
momenta of outgoing quark and anti-quark, respectively. In the above vertices 
we neglect the scalar and pseudoscalar couplings, $k_{\mu}$ and
$k_{\mu}\gamma_5$ with $k=p_q+p_{\bar q}$, since in $e^+e^-$
collisions these terms give contributions proportional
to the electron mass. We note that some of these neglected terms are
needed to maintain the electromagnetic gauge invariance for the axial
vector couplings in Eq.(\ref{ver}).
The vector and axial-vector couplings $V^{Z,\gamma}_q$ and $A^{Z}_q$
contain both the SM and new physics contributions, while $A^{\gamma}_q$
and $S^{Z,\gamma}_q$ contain only new physics contributions. The SM can
also contribute to  $A^{\gamma}_q$ and $S^{Z,\gamma}_q$ at loop level, 
but these effects  are very small
and we neglect them in our calculation. One can write
the vector and axial-vector couplings  as
\begin{eqnarray}
V^{Z,\gamma}_q&=&(V^{Z,\gamma}_q)^0+\delta V^{Z,\gamma}_q,\\
A^{Z,\gamma}_q&=&(A^{Z,\gamma}_q)^0+\delta A^{Z,\gamma}_q,
\end{eqnarray}
where $(V^{Z,\gamma}_q)^0$ and $(A^{Z,\gamma}_q)^0$ represent the SM couplings
and $\delta V^{Z,\gamma}_q,\delta A^{Z,\gamma}_q$ the anomalous new physics
contributions. The SM couplings are given by 
\begin{eqnarray}
(V^{\gamma}_q)^0&=&e_q,\\
(A^{\gamma}_q)^0&=&0,\\
(V^Z_q)^0&\equiv&v_q=2I^{3L}_q-4s_W^2e_q,\\
(A^Z_q)^0&\equiv&a_q=2I^{3L}_q,
\end{eqnarray}
where $e_q$ is the electric charge of the quark in unit of $e$ 
and $I^{3L}_q=\pm 1/2$ the weak isospin. 
The new physics contributions $\delta V^{Z,\gamma}_q$ and
$\delta A^{Z,\gamma}_q$ $(q=b,t)$ can be determined from Appendix B;
they are 
\begin{eqnarray}
 \delta V^Z_b&=&\frac{2s_Wc_W}{e}\frac{vm_Z}{\Lambda^2}\left [
   C_{qW}\frac{c_Wk^2}{2vm_Z}
   +(C_{qB}+C_{bB})\frac{s_Wk^2}{vm_Z}
   -C^{(1)}_{\Phi q}-C^{(3)}_{\Phi q}-C_{\Phi b} \right ],\\
 \delta A^Z_b&=&\frac{2s_Wc_W}{e}\frac{vm_Z}{\Lambda^2}\left [ 
   C_{qW}\frac{c_Wk^2}{2vm_Z}
  +(C_{qB}-C_{bB})\frac{s_Wk^2}{vm_Z}
  -C^{(1)}_{\Phi q}-C^{(3)}_{\Phi q}+C_{\Phi b}\right ],\\
 S^Z_b&=&-\frac{8s_Wc_W}{e}\frac{m_b}{\Lambda^2}\frac{v}{\sqrt 2}\left [ 
   C_{Db}\frac{m_Z}{2v}
   -C_{bW\Phi}c_W-2C_{bB\Phi}s_W)\right ],\\
 \delta V^{\gamma}_b&=&\frac{1}{e}\frac{k^2}{2\Lambda^2}\left [ 
   C_{qW}\frac{s_W}{2}-(C_{qB}+C_{bB})c_W
   \right ],\\ 
 \delta A^{\gamma}_b&=&\frac{1}{e}\frac{k^2}{2\Lambda^2}\left [ 
   C_{qW}\frac{s_W}{2}-(C_{qB}-C_{bB})c_W
   \right ],\\ 
 S^{\gamma}_b&=&\frac{2m_b}{e}\frac{\sqrt 2 v}{\Lambda^2}\left (
   C_{bW\Phi}\frac{s_W}{2}-C_{bB\Phi}c_W \right ),\\
 \delta V^Z_t&=&\frac{2s_Wc_W}{e}\frac{vm_Z}{\Lambda^2}\left [ 
-C_{qW}\frac{c_Wk^2}{2vm_Z}+(C_{tB}+C_{qB})\frac{s_Wk^2}{vm_Z} 
\right.\nonumber\\
& &\hspace{2.0cm}\left. -C^{(1)}_{\Phi q}+C^{(3)}_{\Phi q}-C_{t2}
    -(2C_{tB\Phi}s_W-C_{tW\Phi}c_W)2\sqrt 2 \frac{m_t}{m_Z} \right ],\\ 
 \delta A^Z_t&=&\frac{2s_Wc_W}{e}\frac{vm_Z}{\Lambda^2}\left [ 
-C_{qW}\frac{c_Wk^2}{2vm_Z} -(C_{tB}-C_{qB})\frac{s_Wk^2}{vm_Z}
 -C^{(1)}_{\Phi q}+C^{(3)}_{\Phi q}+C_{t2}\right ],\\ 
 S^Z_t&=&\frac{4s_Wc_W}{e}\frac{m_tm_Z}{\Lambda^2}\frac{1}{\sqrt 2}
     \left [ 
 (2C_{tB\Phi}s_W-C_{tW\Phi}c_W)\frac{2v}{m_Z} +C_{Dt} \right ]\\ 
 \delta V^{\gamma}_t&=&\frac{1}{e}\frac{vm_Z}{\Lambda^2}\left [ 
-C_{qW}\frac{s_Wk^2}{4vm_Z}
-(C_{tB}+C_{qB})\frac{c_Wk^2}{2vm_Z}
+(2C_{tB\Phi}c_W+C_{tW\Phi}s_W)\frac{\sqrt 2  m_t}{m_Z} \right ],\\ 
 \delta A^{\gamma}_t&=&\frac{1}{e}\frac{k^2}{\Lambda^2}\left [ 
-C_{qW}\frac{s_W}{4}
 +(C_{tB}-C_{qB})\frac{c_W}{2}
    \right ],\\ 
 S^{\gamma}_t&=&-\frac{2}{e}\frac{m_tv}{\Lambda^2}\frac{1}{\sqrt 2}
 (2C_{tB\Phi}c_W+C_{tW\Phi}s_W). 
\end{eqnarray}

In terms of the vertices given in Eq.(\ref{ver}), 
the observables $R_b$ and $A^b_{FB}$ at LEP I are given by,
to the order $\frac{1}{\Lambda^2}$, 
\begin{eqnarray}\label{Rb}
R_b&=&R_b^{SM}\left[ 1+2\frac{v_b\delta V^Z_b+a_b\delta A^Z_b}{v_b^2+a_b^2}
      (1-R_b^{SM})\right ],\\ \label{AFB}
A_{FB}^b&=&A_{FB}^{SM}\left [1+\frac{v_b\delta A^Z_b+a_b\delta V^Z_b}{a_bv_b}
-2\frac{v_b\delta V^Z_b+a_b\delta A^Z_b}{v_b^2+a_b^2}\right ],
\end{eqnarray}
where we have neglected the bottom quark mass.
Also, in terms of the vertices of Eq.(\ref{ver}), 
the cross section and forward-backward asymmetry for bottom pair 
production at LEP II and top pair production at the NLC are given by 
\begin{eqnarray}
\sigma^0&=&3\beta_q\left \{
(D_{\gamma\gamma}e_e^2e_q^2
+D_{Z\gamma}e_e v_e e_q v_q)\frac{3-\beta_q^2}{2}
+D_{ZZ}(v_e^2+a_e^2)\left[ \frac{3-\beta_q^2}{2}v_q^2+\beta_q^2 a_q^2\right ]
              \right\},\\
\Delta\sigma&=&3\beta_q\left \{
D_{\gamma\gamma}e_e^2 \left[ (3-\beta_q^2)e_q\delta V^{\gamma}_q-\beta_q^2 e_q S^{\gamma}_q
                     \right ]\right.\nonumber\\
& & \hspace{1cm}+D_{ZZ}(v_e^2+a_e^2)\left[(3-\beta_q^2)v_q\delta V^Z_q
+2\beta_q^2 a_q\delta A^Z_q-\beta_q^2 v_q S^Z_q\right ]\nonumber\\
& &\hspace{1cm}\left. +D_{Z\gamma}e_e v_e\left [\frac{3-\beta_q^2}{2}(e_q\delta V^Z_q
 +v_q\delta V^{\gamma}_q)+\beta_q^2 a_q\delta A^{\gamma}_q
 -\frac{\beta_q^2}{2}(e_q S^Z_q +v_q S^{\gamma}_q)\right ]\right \},\\
\frac{\delta A_{FB}}{A_{FB}^0}&=&\frac{D_{Z\gamma}e_e a_e (e_q\delta A^Z_q
 +a_q\delta V^{\gamma}_q+v_q\delta A^{\gamma}_q)
 +4D_{ZZ}v_e a_e (v_q\delta A^Z_q +a_q\delta V^Z_q)}
{D_{Z\gamma}e_e a_e e_q a_q+4D_{ZZ}a_e v_e v_q a_q}
-\frac{\delta \sigma}{\sigma^0},
\end{eqnarray}
where $\beta_q=\sqrt{1-4m_q^2/s}$ is the velocity of the final quarks and
\begin{eqnarray}
D_{\gamma\gamma}&=&\frac{4\pi \alpha^2}{3s},\\
D_{ZZ}&=&\frac{G_F^2}{96\pi}\frac{sm_Z^4}{(s-m_Z^2)^2+(s\Gamma_Z/m_Z)^2},\\
D_{Z\gamma}&=&\frac{G_F\alpha }{3\sqrt 2}\frac{m_Z^2(s-m_Z^2)}
              {(s-m_Z^2)^2+(s\Gamma_Z/m_Z)^2}.
\end{eqnarray}

Including both the SM coupling and new physics contributions,
the $Wt\bar b$ vertex can be written as 
\begin{eqnarray}
\Gamma^{\mu}_{Wt\bar b}&=&-i\frac{g_2}{\sqrt 2}\left [
  \gamma^{\mu}P_L(1+\kappa_1)+\gamma^{\mu}P_R \kappa_2+p_t^{\mu}P_L \kappa_3
 +p_{\bar b}^{\mu}P_L \kappa_4+p_t^{\mu}P_R \kappa_5
+p_{\bar b}^{\mu}P_R \kappa_6\right ],
\end{eqnarray}
where $P_{L,R}\equiv(1\mp \gamma_5)/2$. 
The form factors from new physics
can be determined from Appendix B as
\begin{eqnarray}
\kappa_1&=&\frac{v^2}{\Lambda^2}\left [ 
 C_{tW\Phi}\frac{\sqrt 2 m_t}{g_2 v}+C^{(3)}_{\Phi q}
   -C_{qW}\frac{k^2}{g_2v^2} \right ],\\
\label{rhand}
\kappa_2&=&\frac{v^2}{\Lambda^2}\left [ 
 C_{bW\Phi}\frac{\sqrt 2 m_t}{g_2 v}+\frac{C_{t3}}{2} \right ],\\
\kappa_3&=&\frac{v}{\Lambda^2}\left [ 
 -C_{tW\Phi}\frac{\sqrt 2 }{g_2}-\frac{C_{Dt}}{\sqrt 2}
   +C_{qW}\frac{m_t}{g_2 v} \right ],\\
\kappa_4&=&\frac{v}{\Lambda^2}\left [ 
 C_{tW\Phi}\frac{\sqrt 2 }{g_2}
   +C_{qW}\frac{m_t}{g_2 v} \right ],\\
\kappa_5&=&-\frac{v}{\Lambda^2}
 C_{bW\Phi}\frac{\sqrt 2}{g_2},\\
\kappa_6&=&\frac{v}{\Lambda^2}\left [
\frac{C_{Db}}{\sqrt 2} + C_{bW\Phi}\frac{\sqrt 2}{g_2} \right ].
\end{eqnarray}
Neglecting the bottom quark mass one gets the cross section for the
subprocess $q_i \bar q_j\rightarrow t \bar b$  
\begin{eqnarray}
\hat{\sigma}_0&=&\frac{g^4}{384\pi}\frac{(\hat{s}-m^2_t)^2}{\hat{s}^2(\hat{s}-
m^2_W)^2}[2\hat{s}+m^2_t],\\
\label{tb}
\Delta \hat{\sigma}&=&\frac{g^4}{384\pi}\frac{(\hat{s}-m^2_t)^2}
{\hat{s}^2(\hat{s}-m^2_W)^2}
\left [ 2(2\hat{s}+m^2_t)\kappa_1+(m^2_t-\hat{s})
         m_t(\kappa_3-\kappa_4)\right ]\nonumber\\
&=&\frac{g^4}{384\pi}\frac{(\hat{s}-m^2_t)^2}
{\hat{s}^2(\hat{s}-m^2_W)^2}\frac{1}{\Lambda^2}
\left\{ 2(2\hat{s}+m^2_t)
   [v^2C^{(3)}_{\Phi q}-\frac{\hat s}{g_2}C_{qW}]
\right.\nonumber\\
& &\left.+(\hat{s}-m^2_t)\frac{m_tv}{\sqrt 2}C_{Dt}
+6\hat s \frac{\sqrt 2  m_tv}{g_2}C_{tW\Phi} \right \}.
\end{eqnarray}
The total cross section of single
 top quark production via $q_i \bar q_j\rightarrow t \bar b$ at
 the Fermilab Tevatron which is obtained by
\begin{equation}
\sigma(s)=\sum_{i,j}\int^1_{\tau_0}\frac{d\tau}{\tau}(\frac{1}{s}
\frac{dL_{ij}}{d\tau})(\hat s \hat \sigma_{ij}),
\end{equation}
where $\tau_0=(M_t+M_b)^2/s$, $s$ is the square of center-of-mass energy,
$\hat s=s\tau$ is the square of center-of-mass energy of the subprocess, and
$dL_{ij}/d\tau$ is the parton luminosity given by
\begin{equation}
\frac{dL_{ij}}{d\tau}=\int^1_{\tau} \frac{dx_1}{x_1}[f^A_i(x_1,\mu)
f^B_j(\tau/x_1,\mu)+(A\leftrightarrow B)],
\end{equation}
where  $A$ and $B$ denote the incident hadrons, $i$ and $j$ are the initial 
partons, and $x_1$ and $x_2$  their longitudinal momentum fractions. 
The functions $f^A_i$ and $f^B_j$ are the parton distribution functions.
\vspace{1cm}

\begin{center} 
    {\Large 4. Classifying physics effects}
\end{center}   

In this section, we classify the operators according to their contribution
to the three-particle vertices which are testable at LEP I, II,
the NLC and the Tevatron, i.e., $Wt\bar b$, $Xt\bar t$ and $Xb\bar b$
($X=\gamma, Z,H,g$).  

From Appendix B we can see that  most operators give contributions
to more than one of the three-particle vertices and therefore tests
of these operators are possible when their coupling strengths are constrained
by one of the vertices. 
In Table 1 we summarize the contributions
of these operators to the couplings which can be tested at present or 
future colliders.  The contribution of an operator to a particular vertex is
denoted by an `$\bigotimes$'. Since the operators contribute to different
combinations of observables, we can reclassify them as
\begin{itemize}
\item Class A-1: Contributing to LEP I and LEP II observables, 
                    $\sigma_{t\bar t}$ and $A_{FB}^t$ at the NLC 
                    and $\sigma_{t\bar b}$ at 
                    the Tevatron. They are $O_{qW}$ and 
                    $O^{(3)}_{\Phi q}$.
\item Class A-2: Contributing to LEP I and LEP II observables, and 
                    $\sigma_{t\bar t}$ and $A_{FB}^t$ at the 
                    NLC, but not to  $\sigma_{t\bar b}$ at the Tevatron.
                    They are $O_{qB}$ and $O^{(1)}_{\Phi q}$.
\item Class A-3: Contributing to LEPI and LEP II observables, and 
                    $\sigma_{t\bar b}$ at the Tevatron.
                    They are $O_{Db}$ and $O_{bW\Phi}$.
\item Class A-4: Contributing to LEP I and LEP II observables.
		    They are $O_{bB}$, $O_{\Phi b}$ and $O_{bB\Phi}$. 
\item Class B-1: Contributing to $\sigma_{t\bar t}$ and $A_{FB}^t$ at the 
                 NLC and
                    $\sigma_{t\bar b}$ at the Tevatron. They are
		    $O_{tW\Phi}$ and $O_{Dt}$.
\item Class B-2: Contributing only to $\sigma_{t\bar t}$ and $A_{FB}^t$
                 at the NLC.
                    They are $O_{t2}$, $O_{tB\Phi}$ and $O_{tB}$.
\item Class B-3: Contributing only to $\sigma_{t\bar b}$ at the Tevatron.
                    It contains only $O_{t3}$.
\item Class C-1: Contributing only to couplings $Ht\bar t$ and $Hb\bar b$, 
                    not to any other vertices. They are $O_{t1}$ and $O_{b1}$. 
\item Class C-2: Contributing to the strong interaction
                 sector. They are $O_{tG\Phi}$, $O_{tG}$, $O_{qG}$, 
                 $O_{bG\Phi}$ and $O_{bG}$.
                 These operators only contribute to the
                 strong interactions of third-family quarks and do
                 not contribute to the electroweak interaction at the
                 level of $1/\Lambda^2$.
\end{itemize}
  
In this new classification scheme, Class A operators include a $Z b\bar b$
or $\gamma b\bar b$ interaction and are currently constrained by $R_b$
and $A_{FB}^b$ at LEP I. Class B operators are not constrained by LEP I 
( at least at tree level), but will affect the future collider observables
under consideration. Class C operators affect neither LEP I observables
nor the future collider observables which arise from the electroweak 
interactions at tree level.

\vspace{1cm}
\begin{center} {\Large 5. Numerical examples and discussions}\end{center}   
\vspace{.5cm}

In this section we present numerical analyses for those operators 
which affect $R_b$ and $A_{FB}^b$ at LEP I and observables at future colliders.
They are the Class A operators defined in the preceding section.
We use the analytic formulae given in Sec.3 and use the most recent LEP I 
data on $R_b$ and $A_{FB}^b$ to  constrain the coefficients of the individual
operators in Classes A-1 through A-4,  and then evaluate their possible 
effects on the electroweak observables at LEP II, the Tevatron upgrade and 
the NLC. Operators in Classes B and C are not presently constrained, at least 
at tree level, or they involve the strong interaction sector, and they will 
not be considered here further.
\vspace{.5cm}

{\large 5.1 The effects of $O^{(3)}_{\Phi q}$ and $O_{qW}$}

From the preceding  section we found that the operators of Class A-1 will
affect the most observables. 
Note that in Ref.~[8] the effects of $O_{qW}$ 
on $R_b$ and $\sigma_{t\bar b}$ at the Tevatron have been evaluated.
The present analyses also include this operator, but we will consider 
its effects in LEP II and the NLC as well.
Presently, the experimental data
of $R_b$ and $A^b_{FB}$ are $+1.8\sigma$ and $-1.8\sigma$ away 
from their SM values, respectively[1]. In the analyses below,
we assume a closer agreement with the SM, say both 
$R_b$ and $A^b_{FB}$ are about $1\sigma$ away the SM predictions, 
and examine the consequences.

We note the new physics of Class A-1  yields
\begin{equation}
\delta V^Z_b=\delta A^Z_b=
\frac{4s_Wc_W}{e}\frac{1}{\Lambda^2}\left [
   C_{qW}\frac{c_Wk^2}{4}- 
   C^{(3)}_{\Phi q}\frac{vm_Z}{2}\right ],
\end{equation} 
which we can express in terms of $R_b$ or $A^b_{FB}$.
We get from Eq.(\ref{Rb}) 
\begin{equation}\label{Vz1}
\delta V^Z_b=\delta A^Z_b=
    \frac{R_b^{exp}-R_b^{SM}}{(1-R_b^{SM})R_b^{SM}}
			  \frac{v_b^2+a_b^2}{2(v_b+a_b)},
\end{equation}
or from Eq.(\ref{AFB})
\begin{equation}\label{Vz2}
\delta V^Z_b=\delta A^Z_b=
    \frac{A_{FB}^{exp}-A_{FB}^{SM}}{A_{FB}^{SM}}
	\frac{v_b a_b}{v_b+a_b}
        \frac{v_b^2+a_b^2}{(v_b-a_b)^2},
\end{equation}
where the experimental data and SM values [1] are 
\begin{eqnarray}\label{data1}
R_b^{SM}&=&0.2158,~~R_b^{exp}=0.2178(11),\\ \label{data2}
A^{SM}_{FB}&=&0.1022,~~A_{FB}^{exp}=0.0979(23).
\end{eqnarray}
Since both $v_b$ and $a_b$ are negative,
we find that Eq.(\ref{Vz1}) yields negative values for $\delta V^Z_b$ 
and Eq.(\ref{Vz2}) yields positive values for $\delta V^Z_b$.
This means that any kind of new physics which yields 
$\delta V^Z_b=\delta A^Z_b$, such as $O_{qW}$, 
$O^{(1)}_{\Phi q}$, $O^{(3)}_{\Phi q}$ and $O_{qB}$,
 can not fit both $R_b$ and $A_{FB}$ within the 
$1\sigma$ bounds of the experimental data at the same time.
If the deviations from the SM values as shown in Eq.(\ref{data1})
and Eq.(\ref{data2}) persist, this class of operators will be
ruled out.

Since the error size in $A_{FB}^b$ is larger than that of $R_b^{SM}$,
we estimate the effect of this class of operators by using only 
the $1\sigma$ bound of $R_b$ to set constraints on
the new physics.
We have from Eq.(\ref{Vz1}) and Eq.(\ref{data1}) 
\begin{equation}\label{bound1}
-0.0080<\delta V^Z_b<-0.0023
\end{equation}
Using this bound and assuming the existence of only $O^{(3)}_{\Phi q}$, 
we get the effects on $\sigma_{b\bar b}$ and $A_{FB}^b$ 
at LEP II ($\sqrt s=200$ GeV), $\sigma_{t\bar t}$ 
and $A_{FB}^t$ at the NLC  ($\sqrt s=500$ GeV, $m_t=175$ GeV),
and the single top production rate at Tevatron 
($\sqrt s=2$ TeV, $m_t=175$  GeV) as 
\begin{eqnarray*}
\begin{array}{lll}
{\rm LEP~ II}~ (e^+e^-\rightarrow b\bar b)~~~~~~&
{\rm NLC}~(e^+e^-\rightarrow t\bar t)~~~~~~& 
{\rm Tevatron}~(p\bar p\rightarrow t\bar b+X) \\
 & & \\
0.4{\rm \%}<\frac{\Delta \sigma}{\sigma^0}<1.3{\rm \%}& 
0.1{\rm \%}<\frac{\Delta \sigma}{\sigma^0}<0.3{\rm \%}& 
0.5{\rm \%}<\frac{\Delta \sigma}{\sigma^0}<1.6{\rm \%}\\
 & & \\
0.2{\rm \%}<\frac{\delta A_{FB}}{A_{FB}^0}<0.6{\rm \%}&
0.7{\rm \%}<\frac{\delta A_{FB}}{A_{FB}^0}<2.6{\rm \%},& \\ 
\end{array}
\end{eqnarray*}
which are too small to be observable.
Using the same bound in Eq.(\ref{bound1}) and assuming only the existence 
of $O_{qW}$ we obtain
\begin{eqnarray*}
\begin{array}{lll}
{\rm LEP II}~ (e^+e^-\rightarrow b\bar b)~~~~~~&
{\rm NLC}~(e^+e^-\rightarrow t\bar t)~~~~~~& 
{\rm Tevatron}~(p\bar p\rightarrow t\bar b+X) \\
 & & \\
2.4{\rm \%}<\frac{\Delta \sigma}{\sigma^0}<8.4 {\rm \%}& 
8.6{\rm \%}<\frac{\Delta \sigma}{\sigma^0}<29.8{\rm \%}& 
6.9{\rm \%}<\frac{\Delta \sigma}{\sigma^0}<24.0{\rm \%}\\
 & & \\
0.3{\rm \%}<\frac{\delta A_{FB}}{A_{FB}^0}<1.0{\rm \%}&
16.3{\rm \%}<\frac{\delta A_{FB}}{A_{FB}^0}<56.8{\rm \%}& \\ 
\end{array}
\end{eqnarray*}
where we have used the CTEQ3L parton distribution functions[10] with
$\mu=\sqrt {\hat s}$ for the calculation of the cross section at the
Tevatron. Except for the $A_{FB}^b$ at LEP II, all the other contributions
are sizable.

Let us consider the expected accuracy of the hadron cross section measurements.
At LEP II the cross section for $e^+e^-\rightarrow hadrons$
can be measured with a high accuracy of 0.7\%[11].
Since new physics 
only contributes to the $b\bar b$ production rate and 
$\sigma^0(e^+e^-\rightarrow b\bar b)/\sigma^0(e^+e^-\rightarrow hadrons)
=0.16$, then $\frac{\Delta \sigma}{\sigma^0}(e^+e^-\rightarrow b\bar b)$
can be measured with an accuracy of 4\%, or better when b-tagging is
employed. At the NLC the top quark properties will be tested to high
accuracy and we expect that the top pair production rate there may 
be measurable with an accuracy of a few percent. At the Tevatron a
deviation larger than 16\% from the SM single top production rate 
is expected to be detectable at Run 3 [12].

The above results then show that the operator $O^{(3)}_{\Phi q}$
constrained by $R_b$ has negligibly small effects on $b\bar b$
production at LEP II, $t\bar t$ production at the NLC and single top
production at the Tevatron. On the contrary, the operator  $O_{qW}$
constrained by $R_b$ can cause observable effects at LEP II, the NLC and the  
Tevatron. In other words, if their effects are not observed at future
colliders, $O_{qW}$ is severely constrained, but $O^{(3)}_{\Phi q}$ is not.
We note that the main reason that $O_{qW}$ has large effects at future 
colliders is that it is momentum dependent, and therefore becomes 
enhanced at higher energies. 
\vspace{0.5cm}

{\large 5.2 The effects of $O_{qB}$ and $O^{(1)}_{\Phi q}$ }

 The operators in Class A-2 ($O_{qB}$ and $O^{(1)}_{\Phi q}$) affect
LEP I and  LEP II observables and $\sigma_{t\bar t}$ 
and $A_{FB}^t$ at
the NLC, but not single top production at hadron colliders. 
Note that $O_{qB}$ is momentum
dependent and $O^{(1)}_{\Phi q}$ is momentum independent.
 Like the operators in Class A-1 analyzed above,
they yield $\delta V^Z_b=\delta A^Z_b$. 
Using the bound given in Eq.(\ref{bound1}), we obtain the contribution
of $O_{qB}$
to  $\sigma_{b\bar b}$ and $A_{FB}^b$ at LEP II ($\sqrt s=200$ GeV), 
$\sigma_{t\bar t}$ and $A_{FB}^t$ at NLC ($\sqrt s=500$ GeV, $m_t=175$ GeV) 
as 
\begin{eqnarray*}
\begin{array}{ll}
{\rm LEP II}~ (e^+e^-\rightarrow b\bar b)~~~~~~~~&
 {\rm NLC}~(e^+e^-\rightarrow t\bar t)\\
 &  \\
-0.6{\rm \%}<\frac{\Delta \sigma}{\sigma^0}<-0.2 {\rm \%}& 
16.5{\rm \%}<\frac{\Delta \sigma}{\sigma^0}<57.4{\rm \%}\\
 & \\
  2.9{\rm \%}<\frac{\delta A_{FB}}{A_{FB}^0}<10.0{\rm \%}&
-144.0{\rm \%}<\frac{\delta A_{FB}}{A_{FB}^0}<-41.4{\rm \%}, \\ 
\end{array}
\end{eqnarray*}
and, in the same way, we obtain the contribution of $O^{(1)}_{\Phi q}$ as
\begin{eqnarray*}
\begin{array}{ll}
{\rm LEP II}~ (e^+e^-\rightarrow b\bar b)~~~~~~~&
 {\rm NLC}~(e^+e^-\rightarrow t\bar t) \\
 &  \\
 0.4{\rm \%}<\frac{\Delta \sigma}{\sigma^0}<1.3{\rm \%}& 
-0.3{\rm \%}<\frac{\Delta \sigma}{\sigma^0}<-0.1{\rm \%}\\
 & \\
 0.2{\rm \%}<\frac{\delta A_{FB}}{A_{FB}^0}<0.6 {\rm \%}&
-2.5{\rm \%}<\frac{\delta A_{FB}}{A_{FB}^0}<-0.7{\rm \%}. \\ 
\end{array}
\end{eqnarray*}
Here we see that the effects of $O^{(1)}_{\Phi q}$ are negligibly small, 
but the effects of $O_{qB}$ on $\sigma_{t\bar t}$ and $A_{FB}^t$ at the NLC
can be quite large. As was the case with $O_{qW}$ in the preceding
section, these large effects are primarily due to the momentum dependence
of $O_{qB}$. So the NLC will be a good place to look for the new
physics operator $O_{qB}$. We should again comment that if the values given in 
Eq.(\ref{data1}) and Eq.(\ref{data2}) persist, this class of operators 
and the Class A-1 operators in the preceding subsection will be ruled out.
\vspace{.5cm}

{\large 5.3 The effects of $O_{bB}$, $O_{\Phi b}$ and $O_{bB\Phi}$}

The operators in Class A-4 ($O_{bB}$, $O_{\Phi b}$ and $O_{bB\Phi}$) 
affect $R_b$ and $A_{FB}^b$ at LEP I and 
$\sigma_{b\bar b}$ and $A_{FB}^b$ at LEP II, but not
top pair production at the NLC or single top production at the
Tevatron upgrade.
Since $O_{bB\Phi}$ only appears in  $S^Z_b$ and 
$S^{\gamma}_b$, its contributions to these observables are proportional to 
$m_b$, which to a good approximation can be set to zero in the 
calculations for $b\bar b$ production at LEP I and LEP II. Thus
the contributions of $O_{bB\Phi}$ are negligible and we only need to consider
$O_{bB}$ and $O_{\Phi b}$. We note that $O_{bB}$ is momentum dependent 
while $O_{\Phi b}$ is momentum independent. 
Unlike the case discussed in sub-sections 5.1 and 5.2, the operators
in this class yield 
$\delta V^Z_b=-\delta A^Z_b$. 

 From Eq.(\ref{Rb}) one gets
\begin{equation}\label{vz3}
\delta V^Z_b=-\delta A^Z_b
    =\frac{R_b^{exp}-R_b^{SM}}{(1-R_b^{SM})R_b^{SM}}
			  \frac{v_b^2+a_b^2}{2(v_b-a_b)},
\end{equation}
and from Eq.(\ref{AFB}) one gets
\begin{equation}\label{vz4}
\delta V^Z_b=-\delta A^Z_b
    =\frac{A_{FB}^{exp}-A_{FB}^{SM}}{A_{FB}^{SM}}
	\frac{v_b a_b}{a_b-v_b}
        \frac{v_b^2+a_b^2}{(v_b+a_b)^2},
\end{equation}
using the values in Eqs.(\ref{data1}) and 
(\ref{data2})
we see that both Eqs.(\ref{vz3}) and  (\ref{vz4}) yield positive values
for $\delta V^Z_b$. The bound from Eq.(\ref{vz3}), again assuming
$1\sigma$ deviation, is found to be
\begin{equation}
0.013<\delta V^Z_b<0.044,
\end{equation}
and the bound from Eq.(\ref{vz4}) is 
\begin{equation}
0.023<\delta V^Z_b<0.075.
\end{equation}
We take the overlap of the two
\begin{equation}\label{bound2}
0.023<\delta V^Z_b<0.044,  
\end{equation}
which is required to have the theoretical values of both $R_b$ and
$A_{FB}^b$ to lie within $1\sigma$ of the experimental data.

Considering $O_{bB}$, 
one gets its contribution to $\sigma_{b\bar b}$ 
and $A_{FB}^b$ at LEP II ($\sqrt s=200$ GeV) to be
\begin{eqnarray}
~23.3{\rm \%}&<&\frac{\Delta \sigma}{\sigma^0}(e^+e^-\rightarrow b\bar b)~<~
 							  44.5{\rm \%},\\
-53.9{\rm \%}&<&\frac{\delta A_{FB}}{A_{FB}^0}(e^+e^-\rightarrow b\bar b)~<
						-28.2{\rm \%}. 
\end{eqnarray}
For $O_{\Phi b}$, the contributions to $\sigma_{b\bar b}$ 
and $A_{FB}^b$ at LEP II ($\sqrt s=200$ GeV) are
\begin{eqnarray}
~0.7{\rm \%}&<&\frac{\Delta \sigma}{\sigma^0}(e^+e^-\rightarrow b\bar b)~<~
 							1.3{\rm \%},\\
-3.3{\rm \%}&<&\frac{\delta A_{FB}}{A_{FB}^0}(e^+e^-\rightarrow b\bar b)~<-1.7
                                                        {\rm \%}. 
\end{eqnarray}
So  if only $O_{bB}$ exists, its effects
are likely observable at LEP II even if both $R_b$ and $A_{FB}^b$ 
lie within the $1\sigma$ bounds of the present data.
As with the operators $O_{qW}$ and $O_{qB}$, this is primarily
due to the momentum dependence of $O_{bB}$.
On the contrary, if only $O_{\Phi b}$ exists, there will be no 
observable effects at LEP II.
\vspace{.5cm}

{\large 5.4 The effects of $O_{bW\Phi}$ and $O_{Db}$ }

The operators $O_{bW\Phi}$ and $O_{Db}$ in Class A-3 affect 
LEP I and LEP II as well as single top quark production at the Tevatron
, $\sigma(p\bar p\rightarrow t\bar b+X)$.
Since both of them only appear in  $S^Z_b$ and 
$S^{\gamma}_b$, their contributions to $R_b$ and $A_{FB}^b$ at LEP I
and $\sigma_{b\bar b}$ and $A_{FB}^b$ at LEP II
are proportional to $m_b$ and hence are negligible.
Further,  as Eq.(\ref{tb}) shows, their contributions to
$\sigma(p\bar p\rightarrow t\bar b+X)$ vanish in the approximation of 
neglecting $m_b$. So they are not constrained by these observables at LEP I,
LEP II and the Tevatron. 

However, as Eq.(\ref{rhand}) shows,
 $O_{bW\Phi}$ contributes to the right-handed weak 
charged current, and thus it will
be strictly constrained by the CLEO measurement of $b\rightarrow s\gamma$
[13]. The latest limit is[14]
\begin{equation}
-0.03<\kappa_2=\frac{C_{bW\Phi}}{\Lambda^2} \frac{\sqrt 2 v m_t}{g_2}<0.00.
\end{equation}
Using this bound and keeping the bottom quark mass, we can evaluate 
its contributions to the observables under consideration.
Of course, its contributions must be very small since they are
not only proportional to $m_b$ but also suppressed by the above bound.
For example, with $m_b=5$ GeV its contribution 
to $\sigma_{b\bar b}$ and $A_{FB}^b$ at LEP II ($\sqrt s=200$ GeV) are
found to be
\begin{eqnarray}
-0.2{\rm \%}&<&\frac{\Delta \sigma}{\sigma^0}(e^+e^-\rightarrow b\bar b)~<~
 							0.0{\rm \%},\\
~0.0{\rm \%}&<&\frac{\delta A_{FB}}{A_{FB}^0}(e^+e^-\rightarrow b\bar b)~<~
                                                        0.2{\rm \%}, 
\end{eqnarray}
which, as expected, are negligibly small. 

So the operator $O_{bW\Phi}$, which contributes to the right-handed weak 
charged current of third-family quarks,  can be further constrained, 
although the coefficient $C_{bW\Phi}$ will not be constrained greatly
unless a process can be found in which its contribution is not proportional 
to $m_b$. The operator $O_{Db}$ will also survive since no observables are 
sensitive to it.
\vspace{.5cm}

\begin{center}{\Large 6. Discussions and summary}\end{center}

From the expressions of $\delta V^Z_b$ and $\delta A^Z_b$ one finds
that the new physics constrained by $R_b$ and $A^b_{FB}$ at LEP I 
can be divided into two types: those yielding $\delta V^Z_b=\delta A^Z_b$
and those yielding $\delta V^Z_b=-\delta A^Z_b$. As the above numerical 
calculations show, operators of the first type, including $O_{qW}$,
$O^{(1)}_{\Phi q}$, $O^{(3)}_{\Phi q}$ and $O_{qB}$ in Classes
A-1 and A-2, can not make the theoretical values of both 
$R_b$ and $A_{FB}^b$ to lie 
within the $1\sigma$ bounds of the experimental data 
at the same time. If one uses the $1\sigma$ bound of $R_b$ to set
constraints to this type of new physics, the strict bounds of
Eq.(\ref{bound1}) are obtained. The two operators $O_{qW}$ and
$O_{qB}$, can give rise 
to visible effects at LEP II, the NLC and/or the upgraded Tevatron.
On the contrary, operators of the second type, including $O_{bB}$
and $O_{\Phi b}$ in Class A-4, can make the theoretical values of both
$R_b$ and $A_{FB}$ within the $1\sigma$ bounds of the experimental
data simultaneously, but the bounds Eq.(\ref{bound2}) are not
so strict as the bounds on the operators of the first type.
$O_{bB}$ in the second type of new physics can cause larger effects 
on observables at LEP II, the Tevatron and the NLC.

A common feature of operators with significant effects on LEP II, etc.,
is that they are momentum dependent, as can be seen from Eq.(45)
and Eq.(46). However the suppression of the effect of an operator 
is more complicated. Take the operator $O_{\Phi b}$ as an example. 
Since it is momentum independent, it does not have the enhanced 
effect going from LEP I to LEP II.
Another reason
for its small effects is that $O_{\Phi b}$ only contributes to the
vertex $Zb\bar b$ but not to the vertex $\gamma b\bar b$, and, as
is well-known, the photon exchange channel is dominant in the $b\bar b$
production at LEP II.   

From the above analyses we can say that if the experimental data
of $R_b$ and $A^b_{FB}$, which are now deviating from their SM values
by $1.8\sigma$ and $-1.8\sigma$ respectively, are both upheld and
the deviations are due to the new physics considered here, then the
new physics cannot be the first type, $O_{qW}$,
$O^{(1)}_{\Phi q}$, $O^{(3)}_{\Phi q}$ or $O_{qB}$ alone;
the second type, $O_{bB}$ or $O_{\Phi b}$, must exist.
In such a situation, the existence of $O_{bB}$ will certainly
give rise to observable effects at LEP II while effects of the operator 
$O_{\Phi b}$ will
be unobservable effects. Thus, if no new physics effects
are observed at LEP II, $O_{bB}$ will be ruled out but $O_{\Phi b}$
will not be. Note that in all the numerical
examples presented in this paper, we did not consider the co-existence
of more than two operators at one time. The detailed analyses of their
effects at LEP II in multi-parameter space is under consideration[15]. 

In summary, we have analyzed the effects of the dimension-six 
CP-conserving operators on the observables $R_b$ and $A^b_{FB}$ at LEP I, 
$\sigma(e^+e^-\rightarrow b\bar b)$ and $A^b_{FB}$ at LEP II,
$\sigma(e^+e^-\rightarrow t\bar t)$ and $A_{FB}$ at NLC as well as 
$\sigma(p\bar p\rightarrow t\bar b+X)$ at the Tevatron. We found that
in the region allowed by $R_b$ and $A^b_{FB}$ at LEP I, some operators
can still have significant contribution to observables at LEP II,
the Tevatron and the NLC, while some other operators have negligibly
small effects and thus can be safely ignored.
\vspace{1cm}

\begin{center}{\Large  Acknowledgement}\end{center}

This work was supported in part by the U.S. Department of Energy, Division
of High Energy Physics, under Grant No. DE-FG02-94ER40817.

\vspace{.5cm}
\appendix{\Large ~~~CP-conserving operators after symmetry breaking}
\vspace{.5cm}

In order to shorten some of the expressions we will use the 
following notations:
\begin{eqnarray} 
W^3_{\mu}&=&-Z_{\mu}\cos\theta_W+A_{\mu}\sin\theta_W,\\
B_{\mu}&=&Z_{\mu}\sin\theta_W+A_{\mu}\cos\theta_W,\\
B_{\mu\nu}&=&Z_{\mu\nu}\sin\theta_W+A_{\mu\nu}\cos\theta_W,\\
W^{\pm,3}_{\mu\nu}&=&\partial_{\mu}W^{\pm,3}_{\nu}
                     -\partial_{\nu}W^{\pm,3}_{\mu},\\
g_Z&=&\frac{2m_Z}{v}=\sqrt {g_1^2+g_2^2} 
\end{eqnarray}

The CP-conserving operators after electroweak symmetry breaking are given as

(1) Class 1
\begin{eqnarray}
O_{t1}&=&\frac{1}{2\sqrt 2}H(H+2v)(H+v)(\bar t t),\\
O_{t2}&=&\frac{1}{2}g_Z(H+v)^2 Z^{\mu}(\bar t_R \gamma_{\mu} t_R),\\
O_{t3}&=&\frac{1}{2\sqrt 2}g_2 (H+v)^2\left [
    W_{\mu}^+ (\bar t_R\gamma^{\mu} b_R)
    +W_{\mu}^- (\bar b_R\gamma^{\mu} t_R)\right ],\\
O_{Dt}&=&\frac{1}{2\sqrt 2}\partial^{\mu}H \left [\partial_{\mu}(\bar t t)
    +\bar t\gamma_5\partial_{\mu}t-(\partial_{\mu}\bar t)\gamma_5 t
    -i\frac{4}{3}g_1 B_{\mu}\bar t \gamma_5 t\right ]\nonumber\\
& & +i\frac{1}{4\sqrt 2}g_Z (H+v)Z^{\mu}\left [\bar t\partial_{\mu}t
    -(\partial_{\mu}\bar t) t+\partial_{\mu}(\bar t\gamma_5 t)
-i\frac{4}{3}g_1 B_{\mu}\bar t t\right ]\nonumber\\
& & -i\frac{1}{2}g_2 (H+v)W_{\mu}^- \left [\bar b_L\partial^{\mu} t_R
                 -i\frac{2}{3}g_1 B^{\mu}\bar b_L  t_R\right ]\nonumber\\
& & +i\frac{1}{2}g_2 (H+v)W_{\mu}^+ \left [(\partial^{\mu} \bar t_R)b_L
                 +i\frac{2}{3}g_1 B^{\mu}\bar t_R  b_L\right ],\\
O_{tW\Phi}&=&\frac{1}{2\sqrt 2}(H+v)(\bar t\sigma^{\mu\nu}t)
\left [W^3_{\mu\nu}-ig_2(W^+_{\mu}W^-_{\nu}-W^-_{\mu}W^+_{\nu})\right ]\nonumber\\
& & +\frac{1}{2}(H+v)(\bar b_L\sigma^{\mu\nu} t_R)
\left [W^-_{\mu\nu}-ig_2(W^-_{\mu}W^3_{\nu}-W^3_{\mu}W^-_{\nu})\right ]\nonumber\\
& & +\frac{1}{2}(H+v)(\bar t_R\sigma^{\mu\nu} b_L)
\left [W^+_{\mu\nu}-ig_2(W^3_{\mu}W^+_{\nu}-W^+_{\mu}W^3_{\nu})\right ],\\
O_{tB\Phi}&=&\frac{1}{\sqrt 2}(H+v)(\bar t\sigma^{\mu\nu}t)B_{\mu\nu},\\
O_{tB}&=&\left [\bar t_R\gamma^{\mu} \partial^{\nu}t_R
         +\partial^{\nu} \bar t_R\gamma^{\mu}t_R\right ]B_{\mu\nu},\\
O_{tG\Phi}&=&\frac{1}{\sqrt 2}(H+v)(\bar t\sigma^{\mu\nu}T^A t)
				G^A_{\mu\nu},\\
O_{tG}&=&\left [\bar t_R\gamma^{\mu}T^A \partial^{\nu}t_R
+\partial^{\nu} \bar t_R\gamma^{\mu}T^A t_R\right ]G^A_{\mu\nu}
+ig_s \bar t_R\gamma^{\mu}\left [G^{\nu},G_{\mu\nu}\right ]t_R,
\end{eqnarray}
(2) Class 2
\begin{eqnarray}
O_{qG}&=&\left [\bar q_L \gamma^{\mu}T^A \partial^{\nu}q_L
         +\partial^{\nu}\bar q_L \gamma^{\mu}T^A q_L\right ]
          G^A_{\mu\nu}
  +ig_s \bar q_L\gamma^{\mu}\left\{G^{\nu},G_{\mu\nu}\right\}q_L,\\
O_{qW}&=&\frac{1}{2}W^3_{\mu\nu}\left [\bar t_L\gamma^{\mu}\partial^{\nu}t_L
+\partial^{\nu}\bar t_L\gamma^{\mu}t_L
-\bar b_L\gamma^{\mu}\partial^{\nu}b_L
-\partial^{\nu}\bar b_L\gamma^{\mu}b_L\right ]  \nonumber\\
& & +\frac{1}{\sqrt 2}\left [W^+_{\mu\nu}(\bar t_L\gamma^{\mu}\partial^{\nu}b_L
+\partial^{\nu}\bar t_L\gamma^{\mu}b_L)
+W^-_{\mu\nu}(\bar b_L\gamma^{\mu}\partial^{\nu}t_L
+\partial^{\nu}\bar b_L\gamma^{\mu}t_L) \right ]\nonumber\\
& & -ig_2\bar q_L \gamma^{\mu} \left [W_{\mu},W_{\nu}\right ]\partial^{\nu}q_L
  -ig_2\partial^{\nu}\bar q_L \gamma^{\mu} 
  \left [W_{\mu},W_{\nu}\right ]q_L
  -ig_2 \bar q_L \gamma^{\mu} \left [W_{\mu\nu},W^{\nu}\right ]q_L,\\
O_{qB}&=&B_{\mu\nu}\left [\bar q_L\gamma^{\mu} \partial^{\nu}q_L
         +\partial^{\nu} \bar q_L\gamma^{\mu}q_L\right ],\\
O_{bG}&=&\left [\bar b_R\gamma^{\mu}T^A \partial^{\nu}b_R
         +\partial^{\nu} \bar b_R\gamma^{\mu}T^A b_R\right ]G^A_{\mu\nu}
         -ig_s \bar b_R\gamma_{\mu}\left [G^{\mu\nu},G_{\nu}\right ]b_R,\\
O_{bB}&=&\left [\bar b_R\gamma_{\mu}\partial_{\nu}b_R
         +\partial_{\nu} \bar b_R\gamma_{\mu} b_R\right ]B^{\mu\nu},\\
O_{\Phi q}^{(1)}&=&\frac{1}{2}g_Z(H+v)^2Z_{\mu}\left [\bar t_L\gamma^{\mu}t_L
          +\bar b_L\gamma^{\mu}b_L\right ] ,\\
O_{\Phi q}^{(3)}&=&-\frac{1}{2}g_Z(H+v)^2Z_{\mu}\left [\bar t_L\gamma^{\mu}t_L
          -\bar b_L\gamma^{\mu}b_L\right ]\nonumber\\
& &    +\frac {1}{\sqrt 2}g_2(H+v)^2\left [W^+_{\mu}\bar t_L\gamma^{\mu}b_L
		+W^-_{\mu}\bar b_L\gamma^{\mu}t_L\right ],\\
O_{\Phi b}&=&\frac{1}{2}g_Z(H+v)^2Z_{\mu} \bar b_R \gamma^{\mu}b_R,\\
O_{b1}&=&\frac{1}{2\sqrt 2}H(H+v)(H+2v) \bar b  b,\\
O_{Db}&=&\frac{1}{2\sqrt 2}\partial^{\mu}H \left [\partial_{\mu}(\bar b b)
  +\bar b\gamma_5\partial_{\mu}b-(\partial_{\mu}\bar b)\gamma_5 b
   +\frac{2}{3}g_1B_{\mu}\bar bi\gamma_5  b\right ]\nonumber\\
& & +\frac{i}{4\sqrt 2}g_Z (H+v)Z^{\mu}\left [(\partial_{\mu}\bar b) b
     -\bar b\partial_{\mu}b-\partial_{\mu}(\bar b \gamma_5b)
-i\frac{2}{3}g_1B_{\mu}(\bar b b)\right ]\nonumber\\
& & -\frac{i}{2}g_2 (H+v)\left [W_{\mu}^+ (\bar t_L \partial^{\mu}b_R
                 +i\frac{g_1}{3}B^{\mu} \bar t_L  b_R)
-W_{\mu}^- ( \partial^{\mu}\bar b_R t_L
-i\frac{g_1}{3}B^{\mu} \bar b_R t_L)\right ],\\
O_{bW\Phi}&=&\frac{1}{2}(H+v)\left [W^+_{\mu\nu}(\bar t_L \sigma^{\mu\nu} b_R)
	+W^-_{\mu\nu}(\bar b_R \sigma^{\mu\nu} t_L)
-\frac{1}{\sqrt 2}W^3_{\mu\nu}(\bar b \sigma^{\mu\nu} b)\right.\nonumber\\
& & +ig_2(W^+_{\mu}W^3_{\nu}-W^3_{\mu}W^+_{\nu})(\bar t_L \sigma^{\mu\nu} b_R)
    -ig_2(W^-_{\mu}W^3_{\nu}-W^3_{\mu}W^-_{\nu})(\bar b_R \sigma^{\mu\nu} t_L)
\nonumber\\
& &\left.+i\frac{g_2}{\sqrt 2}
(W^+_{\mu}W^-_{\nu}-W^-_{\mu}W^+_{\nu})(\bar b \sigma^{\mu\nu}b)\right ],\\    
O_{bB\Phi}&=&\frac{1}{\sqrt 2}(H+v) B_{\mu\nu}
            (\bar b \sigma^{\mu\nu} b),\\
O_{bG\Phi}&=&\frac{1}{\sqrt 2}(H+v) G^A_{\mu\nu}
            (\bar b \sigma^{\mu\nu} T^A b)
\end{eqnarray}

\vspace{.5cm}
\appendix{\Large~~~Effective Lagrangian for some couplings}
\vspace{.5cm}

The effective Lagrangian for the couplings $Wt\bar b$, $Xb\bar b$
and $Xt\bar t$ $(X=Z,\gamma,g,H)$ are given by (the SM
Lagrangians are not included here)

\begin{eqnarray}
{\cal L}_{Wt\bar b}&=&\frac{C_{\Phi q}^{(3)}}{\Lambda^2}\frac {g_2}{\sqrt 2}v^2
      W^+_{\mu}(\bar t\gamma^{\mu}P_L b)
+ \frac{C_{t3}}{\Lambda^2} \frac{v^2}{2}\frac{g_2}{\sqrt 2} 
               W_{\mu}^+ (\bar t\gamma^{\mu}P_R b)\nonumber\\
& &+\frac{C_{Dt}}{\Lambda^2} \frac{v}{\sqrt 2}\frac{g_2}{\sqrt 2}
     W_{\mu}^+ (i\partial^{\mu} \bar t)P_L b
     -\frac{C_{Db}}{\Lambda^2}\frac{v}{\sqrt 2}\frac{g_2}{\sqrt 2} 
      W_{\mu}^+ (i\bar t P_R \partial^{\mu}b)\nonumber\\
& &+\frac{C_{tW\Phi}}{\Lambda^2}\frac{v}{2}
        W^+_{\mu\nu}(\bar t\sigma^{\mu\nu}P_L b)
+\frac{C_{bW\Phi}}{\Lambda^2}\frac{v}{2} 
      W^+_{\mu\nu}(\bar t \sigma^{\mu\nu}P_R b)\nonumber\\
& & +\frac{C_{qW}}{\Lambda^2} \frac{1}{\sqrt 2}
        W^+_{\mu\nu}(\bar t\gamma^{\mu}P_L\partial^{\nu}b
                     +\partial^{\nu}\bar t\gamma^{\mu}P_L b),\\
{\cal L}_{Zb\bar b}&=&(\frac{C_{\Phi q}^{(1)}}{\Lambda^2}
  +\frac{C_{\Phi q}^{(3)}}{\Lambda^2})(v m_Z) 
    Z_{\mu}(\bar b\gamma^{\mu}P_L b)
  +\frac{C_{\Phi b}}{\Lambda^2}(v m_Z) 
    Z_{\mu}(\bar b \gamma^{\mu}P_R b)\nonumber\\
& & +(\frac{C_{qW}}{\Lambda^2}\frac{c_W}{2}
                  +\frac{C_{qB}}{\Lambda^2}s_W) 
   Z_{\mu\nu}(\bar b \gamma^{\mu}P_L \partial^{\nu}b
                      +\partial^{\nu}\bar b \gamma^{\mu}P_L b)\nonumber\\
& &+\frac{C_{bB}}{\Lambda^2}s_W
    Z_{\mu\nu}(\bar b \gamma^{\mu}P_R \partial^{\nu}b
                +\partial^{\nu} \bar b\gamma^{\mu}P_R b)\nonumber\\
& &+\frac{m_Z}{2\sqrt 2}  
    Z^{\mu}\left [i(\partial_{\mu}\bar b b-\bar b\partial_{\mu}b)
                 \frac{C_{Db}}{\Lambda^2}
            -i\partial_{\mu}(\bar b \gamma_5b)
\frac{C_{Db}}{\Lambda^2}\right ]\nonumber\\
& &+(\frac{C_{bW\Phi}}{\Lambda^2}\frac{c_W}{2}
+\frac{C_{bB\Phi}}{\Lambda^2}s_W)\frac{v}{\sqrt 2} 
   Z_{\mu\nu}(\bar b \sigma^{\mu\nu} b),\\
{\cal L}_{\gamma b\bar b}&=&
  (\frac{C_{qB}}{\Lambda^2}c_W-\frac{C_{qW}}{\Lambda^2}\frac{s_W}{2})
  A_{\mu\nu}(\bar b \gamma^{\mu}P_L \partial^{\nu}b
           +\partial^{\nu}\bar b \gamma^{\mu}P_L b)\nonumber\\
& &+\frac{C_{bB}}{\Lambda^2}c_W
   A_{\mu\nu}(\bar b \gamma^{\mu}P_R \partial^{\nu}b
         +\partial^{\nu} \bar b\gamma^{\mu}P_R b)\nonumber\\
& &+(\frac{C_{bB\Phi}}{\Lambda^2}c_W
     -\frac{C_{bW\Phi}}{\Lambda^2}\frac{s_W}{2})\frac{v}{\sqrt 2}
   A_{\mu\nu}(\bar b \sigma^{\mu\nu} b),\\
{\cal L}_{Zt\bar t}&=&
(\frac{C_{\Phi q}^{(1)}}{\Lambda^2}-\frac{C_{\Phi q}^{(3)}}{\Lambda^2})
    v m_ZZ_{\mu}(\bar t\gamma^{\mu}P_Lt)
+\frac{C_{t2}}{\Lambda^2} v m_Z Z^{\mu}
                 (\bar t \gamma_{\mu}P_R t)\nonumber\\
& &+\frac{C_{Dt}}{\Lambda^2}\frac{m_Z}{2\sqrt 2}
    Z^{\mu}\left [i\bar t\partial_{\mu}t-i(\partial_{\mu}\bar t) t
                  +i\partial_{\mu}(\bar t\gamma_5 t)\right ]\nonumber\\
& &+(\frac{C_{tB\Phi}}{\Lambda^2}s_W
 -\frac{C_{tW\Phi}}{\Lambda^2}\frac{c_W}{2})\frac{v}{\sqrt 2}
      Z_{\mu\nu}(\bar t\sigma^{\mu\nu}t)
   +\frac{C_{tB}}{\Lambda^2}s_W
      Z_{\mu\nu}(\bar t\gamma^{\mu}P_R \partial^{\nu}t
         +\partial^{\nu} \bar t\gamma^{\mu}P_Rt)\nonumber\\
& &+(\frac{C_{qB}}{\Lambda^2}s_W-\frac{C_{qW}}{\Lambda^2}\frac{c_W}{2}) 
      Z_{\mu\nu}(\bar t \gamma^{\mu}P_L \partial^{\nu}t
           +\partial^{\nu}\bar t \gamma^{\mu}P_L t),\\
{\cal L}_{\gamma t\bar t}&=&(\frac{C_{tW\Phi}}{\Lambda^2}\frac{s_W}{2}
   +\frac{C_{tB\Phi}}{\Lambda^2}c_W)\frac{v}{\sqrt 2}
A_{\mu\nu}(\bar t\sigma^{\mu\nu}t)
  +\frac{C_{tB}}{\Lambda^2}c_W
   A_{\mu\nu}(\bar t\gamma^{\mu}P_R \partial^{\nu}t
         +\partial^{\nu} \bar t\gamma^{\mu}P_Rt)\nonumber\\
& &
  +(\frac{C_{qB}}{\Lambda^2}c_W+\frac{C_{qW}}{\Lambda^2}\frac{s_W}{2})
        A_{\mu\nu}(\bar t \gamma^{\mu}P_L \partial^{\nu}t
         +\partial^{\nu} \bar t \gamma^{\mu}P_L t),\\
{\cal L}_{H t\bar t}&=&\frac{C_{t1}}{\Lambda^2} \frac{v^2}{\sqrt 2}H(\bar t t)
+\frac{C_{Dt}}{\Lambda^2}
\frac{1}{2\sqrt 2}\partial^{\mu}H 
    \left [\partial_{\mu}(\bar t t)
    +\bar t\gamma_5\partial_{\mu}t-(\partial_{\mu}\bar t)\gamma_5 t
    \right ],\\
{\cal L}_{g t\bar t}&=&\frac{C_{tG}}{\Lambda^2}
  \left [\bar t\gamma^{\mu}P_RT^A \partial^{\nu}t
         +\partial^{\nu} \bar t\gamma^{\mu}P_R T^A t \right ]G^A_{\mu\nu}
\nonumber\\
& & +\frac{C_{qG}}{\Lambda^2}
\left [\bar t\gamma^{\mu}P_L T^A \partial^{\nu}t
         +\partial^{\nu}\bar t \gamma^{\mu}P_L T^A t\right ] G^A_{\mu\nu} 
+\frac{C_{tG\Phi}}{\Lambda^2}
  \frac{v}{\sqrt 2}(\bar t\sigma^{\mu\nu}T^A t)G^A_{\mu\nu},\\
{\cal L}_{Hb\bar b}&=&
\frac{C_{b1}}{\Lambda^2}\frac{v^2}{\sqrt 2}H(\bar b b)
+\frac{C_{Db}}{\Lambda^2}
   \frac{1}{2\sqrt 2}\partial^{\mu}H \left [
   \bar b\gamma_5\partial_{\mu}b-(\partial_{\mu}\bar b)\gamma_5b
+\partial_{\mu}(\bar b b)\right ].
\end{eqnarray}
\vspace{3cm}

{\LARGE References}
\vspace{0.3in}
\begin{itemize}
\begin{description}
\item[{\rm [1]}] For a recent review, see, 
                  G.Altarelli, CERN-TH/96-265, hep-ph/9611239.
\item[{\rm [2]}]CDF Collaboration,  Phys.Rev.Lett. {\bf 74}, 2626(1995);\\
		D0 Collaboration,  Phys.Rev.Lett. {\bf74}, 2632(1995).
\item[{\rm [3]}] C.J.C.Burgess and H.J.Schnitzer, Nucl.Phys.B228, 454(1983);\\
		 C.N.Leung, S.T.Love and S.Rao, Z.Phys.C31, 433(1986);\\
		 W.Buchmuller and D.Wyler, Nucl.Phys.B268, 621(1986).
\item[{\rm [4]}] For example, see, K.Hagiwara et al, Phys.Lett.B283, 353(1992); 
                                   Phys. Rev. D48, 2182 (1993).
\item[{\rm [5]}] G.J.Gounaris, F.M.Renard and C.Verzegnassi, Phys.Rev.D52
                                                             451(1995).
\item[{\rm [6]}] G.J.Gounaris, D.T.Papadamou and F.M.Renard,
                 hep-ph/9609437.
\item[{\rm [7]}] R.D. Peccei and X. Zhang, Nucl. Phys. B337, 269 (1990);\\
               R.D. Peccei, S. Peris and X. Zhang, Nucl. 
               Phys. B349, 305 (1991);\\
              C.T. Hill and S. Parke, Phys. Rev. D49, 4454 (1994);\\
              D. Atwood, A. Kagan and T. Rizzo, Phys. Rev. D52, 6264 (1995); \\
             D.O.Carlson, E.Malkawi and C.-P.Yuan, Phys.Lett.B337,145(1994);\\
         H. Georgi, L. Kaplan, D. Morin and A. Shenk, Phys. Rev. D51, 3888
          (1995);\\ 
	     T.Han, R.D.Peccei and X. Zhang, Nucl.Phys.B454, 527(1995);\\
             X.Zhang and B.-L.Young,  Phys.Rev.D51, 6564(1995);\\
             E. Malkawi and T. Tait, Michigan State Preprint, MSUHEP-51116,
               (November 1995);\\
            S. Dawson and G. Valencia, Phys. Rev. D53, 1721 (1996);\\
             T.G.Rizzo,  Phys.Rev.D53, 6218(1996);\\
	     P.Haberl, O.Nachtman and A.Wilch,  Phys.Rev.D53, 4875(1996);\\
      T.Han, K.Whisnant and B.-L.Young and X.Zhang, Phys.Lett.B385, 311(1996);\\
      T.Han, K.Whisnant and B.-L.Young and X.Zhang, hep-ph/9603247;\\
      K.Hagiwara, T.Hatsukano, S.Ishihara and R.Szalapski, hep-ph/9612268;\\
      G.J.Gounaris, J.Layssac and F.M.Renard, hep-ph/9612335.
\item[{\rm [8]}]  A.Data and X.Zhang, hep-ph/9611247.
\item[{\rm [9]}]  B.-L.Young, hep-ph/9511282.
\item[{\rm [10]}] H.L. Lai et.al., Phys.Rev.D51, 4763(1995).
\item[{\rm [11]}] F.Boudjema, B.Mele et al., in: {\it Physics at LEP 2},
                  CERN-96-01, eds. G.Altarelli, T. Sjostrand, F.Zwirner.
\item[{\rm [12]}] A.P.Heinson, ``Future Top Physics at the Tevatron and LHC",
                  hep-ph/9605010(1996).
\item[{\rm [13]}] M.Alam et al., CLEO Collaboration, Phys.Rev.Lett.74, 
                                                          2885(1995).
\item[{\rm [14]}] M.Hosch, K.Whisnant and B.-L.Young, hep-ph/9607413, to be 
                  published in Phys.Rev.D.
\item[{\rm [15]}] A.Data, K.Whisnant, B.-L.Young and X.Zhang, work in progress.
\end{description}
\end{itemize}
\eject

~~~\\
Table 1:\\
Dimension-six CP-conserving operators contributive to the vertice which
can be tested at present or future colliders. 
The contribution of an operator to a particular vertex is
denoted by an `$\bigotimes$'.
Operators with significant observable effects at LEP II, the Tevatron and 
the NLC are marked by `$\star$'.
\vspace{0.1in}

\begin{center}
\begin{tabular}{|l|l|c|c|c|c|c|c|c|c|c|}
\hline
 \multicolumn{2}{|c|}{ } &\multicolumn{2}{c|}{ } & &\multicolumn{2}{c|}{ }
                  &\multicolumn{2}{c|}{}&\multicolumn{2}{c|}{}\\
 \multicolumn{2}{|c|}{ } 
   &\multicolumn{2}{c|}{$R_b$, $\sigma_{b\bar b}$, $A_{FB}^b$}
   &$\sigma_{t\bar b}$ 
   &\multicolumn{2}{c|}{$\sigma_{t\bar t}$, $A_{FB}^t$}
  &\multicolumn{2}{c|}{Strong}&\multicolumn{2}{c|}{Yukawa}\\
 \multicolumn{2}{|c|}{ } &\multicolumn{2}{c|}{LEP I, II}
   &Tevatron 
   &\multicolumn{2}{c|}{ NLC}
   &\multicolumn{2}{c|}{couplings}&\multicolumn{2}{c|}{couplings}\\
 \multicolumn{2}{|c|}{ } &\multicolumn{2}{c|}{ } & &\multicolumn{2}{c|}{ }
                     &\multicolumn{2}{c|}{ } &\multicolumn{2}{c|}{ } \\ \hline
 \multicolumn{2}{|c|}{ }& & & & & & & & & \\ 
 \multicolumn{2}{|c|}{ }& $~Zb\bar b~$ & $~\gamma b\bar b~$
   &$~Wt\bar b~$  & $~Zt\bar t~$ & $~\gamma t\bar t~$ 
   & $~gt\bar t~$ & $~gb\bar b~$
   & $~Ht\bar t~$ & $~Hb\bar b~$  \\ 
 \multicolumn{2}{|c|}{ }& & & & & & & & & \\ \hline
$A-1$ & $~~~O_{qW}~\star~$ 
    & $\bigotimes$ & $\bigotimes$ & $\bigotimes$ 
    & $\bigotimes$ & $\bigotimes$ & & & & \\ \cline{2-11}
    & $~~~O_{\Phi q}^{(3)}$ 
    & $\bigotimes$ & &$\bigotimes$&$\bigotimes$ & & & & &  \\ \hline
$A-2$ & $~~~O_{qB}~\star~$ 
    &$\bigotimes$&$\bigotimes$&  
    & $\bigotimes$ & $\bigotimes$ & & & & \\ \cline{2-11}
    & $~~~O_{\Phi q}^{(1)}$
    &$\bigotimes$ & & &$\bigotimes$& & & & & \\ \hline
$A-3$ & $~~~O_{Db}$
    &$\bigotimes$ & & $\bigotimes$& & & & & &$\bigotimes$ \\ \cline{2-11}
    & $~~~O_{bW\Phi}$
    &$\bigotimes$&$\bigotimes$ & $\bigotimes$& & & & & &  \\ \hline
$A-4$ & $~~~O_{bB}~\star~$
    &$\bigotimes$&$\bigotimes$& & & & & & & \\ \cline{2-11}
    & $~~~O_{\Phi b}$
    &$\bigotimes$ & & & & & & & &  \\ \cline{2-11}
    & $~~~O_{bB\Phi}$
    & $\bigotimes$&$\bigotimes$ & & & & & & & \\ \hline   
$B-1$ & $~~~O_{Dt}$
    & & & $\bigotimes$&$\bigotimes$& & & &$\bigotimes$& \\ \cline{2-11}
    & $~~~O_{tW\Phi}$
    & & & $\bigotimes$&$\bigotimes$&$\bigotimes$& & & & \\ \hline
$B-2$ & $~~~O_{tB}$
    & & & &$\bigotimes$&$\bigotimes$& & & & \\ \cline{2-11}
    & $~~~O_{tB\Phi}$
    & & & &$\bigotimes$&$\bigotimes$& & & & \\ \cline{2-11}
    & $~~~O_{t2}$
    & & & &$\bigotimes$& & & & &  \\ \hline
$B-3$ & $~~~O_{t3}$& & &$\bigotimes$& & & & & & \\ \hline
$C-1$ & $~~~O_{t1}$& & & & & & & &$\bigotimes$ & \\ \cline{2-11}
    & $~~~O_{b1}$& & & & & & & & &$\bigotimes$  \\ \hline
$C-2$ & $~~~O_{tG}$& & & & & & $\bigotimes$ & & & \\ \cline{2-11}
    & $~~~O_{qG}$& & & & & & $\bigotimes$ & $\bigotimes$ & & \\ \cline{2-11}
    & $~~~O_{bG}$& & & & & & & $\bigotimes$ & & \\ \cline{2-11}
    & $~~~O_{tG\Phi}$& & & & & & $\bigotimes$ & & & \\ \cline{2-11}
    & $~~~O_{bG\Phi}$& & & & & & &$\bigotimes$ & & \\
    & & & & & & & & & & \\ 
\hline
\end{tabular}
\end{center}
\end{document}